\documentclass[12pt]{article}
\begin{document}
\bibliographystyle{unsrt}
\title {Propagator for a  spin 1/2 particle in terms of  the unitary representations of the Lorentz group}  
\date{\today}
\author{R. A. Frick  \thanks{Email: rf@thp.uni-koeln.de}\\{Institute for Theoretical Physics,}\\{ University of Cologne, D-50923 K\"oln , Germany}}
\date{\today}
\maketitle
\begin{abstract}
In this paper we extend our previous result on the description of the
partcle motion in a generalized Heisenberg picture to a relativistic
fermion.  The operators of the Lorentz algebra in this picture may be
regarded as field operators. In this approach the transition
amplitudes for the particle are constructed in terms of two-component
functions in the unitary representations of the Lorentz group.
\end{abstract}

 In \cite{Frick1} it was found that the propagation  of a massive relativistic particle  may be defined as  space-time transition between states with equal eigenvalues of the first  and  second Casimir operators $C_1$ and $C_2$ of the Lorentz algebra  in the unitary representations . In addition a vector on the light-cone $n$  $(n^2_0-{\bf n}^2=0)$ was used. An integral representation for  the transition amplitude for a massive particle with spin 0 has been obtained.
This  method for describing  the particle motion is based on a  generalized Heisenberg/Schr\"odinger  picture in which either the analogue of Heisenberg states or the analogue of Schr\"odinger operators are independent of both time and space coordinates t, ${\bf x}$ \cite{Frick2}. There is no ${\bf x}$ representation.

In the present work  we   apply   this approach for such a practically important example as spin 1/2 particle. The unitary  representations of the Lorentz group correspond to the eigenvalues $1+\alpha^2-{\lambda}^2$  of the operator $C_1$  and the eigenvalues $\alpha\lambda$ of the operator $C_2$ $(0\leq\alpha<\infty,\quad\lambda=-s,...,s; s=spin)$ \cite{Joos}. We find for $\lambda=-1/2,1/2$ the eigenfunctions of both operators $C_1$ and $C_2$ which contain the vector on the light-cone $n$ and use these functions to construct  the transition amplitudes for a spin 1/2 particle.  

At first for the subsequent presentation we will give a short review for the description of the particle motion   in  the generalized Heisenberg  picture. In this picture   the coordinates t, ${\bf x}$ occur equal in the description and the operators of the Lorentz algebra  (${\bf N}$, ${\bf J}$ - are  space-time independent operators)
\begin{equation}
\label{1}
{\bf N}(x)=S^{-1}(x){\bf N}S(x)={\bf N}+t{\bf P}-{\bf x}{H},
\end{equation} 
\begin{equation}
\label{2}
{\bf J}(x)=S^{-1}(x){\bf J}S(x)={\bf J}-{\bf x}\times{\bf P},
\end{equation}
may be  considered as field  operators. In (\ref{1}), (\ref{2}) ${H}$ and ${\bf P}$ are the  Hamilton and momentum operators of the particle in the generalized Schr\"odinger  picture and $S(x)=\exp[-i(tH-{\bf x}\cdot{\bf P})]$.   The  equations for this field may be written in the form in which the  operators ${H}$,  ${\bf P}$  play the source role 
\begin{equation}
\label{3} 
{\bf \nabla}_{\bf x}\times{\bf J}(x)=\frac{\partial{\bf N}
(x)}{\partial{t}}+{\bf P},\quad{\bf \nabla}_{\bf x}\cdot{\bf J}(x)=0,
\end{equation}
\begin{equation}
\label{4} 
{\bf \nabla}_{\bf x}\cdot{\bf N}(x)=-3H,\quad{\bf \nabla}_{\bf x}\times{\bf N}(x)=0.
\end{equation}
In the classical version of this  field along  the trajectory  of the particle $({\bf x}_t={\bf x}_0+(t-t_0){\bf P}/H)$ for two points  one can find ($C_1(x)={\bf N}^2(x)-{\bf J}^2(x)$, ${C_2}(x)={\bf N}(x)\cdot{\bf J}(x)$)  
\begin{equation}
\label{5} 
{C_1(x_1)}={C_1(x_2)},\quad{C_2}(x_1)={C_2(x_2)}.
\end{equation}
The classical analogue of the operators ${\bf N}$, ${\bf J}$   must be separated from the integrals  of  motion of the particle. In this case the conversion from the  relativistic mechanics to the  quantum version  takes place. This leads  to  the expression for the transition amplitude  of the particle in terms of  the   eigenstates  $|{n},{\lambda},{\alpha}>$ of the operators $C_1$, $C_2$ and the vector ${n}$ 
\begin{eqnarray}
\label{6}
K(x_2;x_1,{\alpha},{\lambda},n)=<{\alpha},{\lambda},{n}|S(x_2-x_1)|{ n}^{'},{\lambda}^{'},{\alpha}^{'}>{_{{\alpha},{\lambda},{ n}={\alpha}^{'},{\lambda}^{'},{ n}^{'}}}.
\end{eqnarray}

We use the momentum representation  (m = mass, $p_0=\sqrt{m^2+{\bf p}^2}$,  ${\vec \sigma}$ - are the Pauli matrices)
\begin{equation}
\label{7}
{\bf N}=ip_0{\nabla}_{\bf p}-\frac{{\vec \sigma}\times{\bf p}}
{2(p_0+m)},\quad{\bf J}=-i{\bf p}\times{\bf \nabla}_{\bf p}
+\frac{\vec \sigma}{2}.                    
\end{equation}
We look for the solutions of the equations 
\begin{equation}
\label{8}
C_1\,\zeta_{\lambda}({\bf p};{\alpha},{\bf n})= (1+\alpha^2-1/4)\,\zeta_{\lambda}({\bf p};{\alpha},{\bf n}),
\end{equation}
\begin{equation}
\label{9}
C_2\,\zeta_{\lambda}({\bf p};{\alpha},{\bf n})={\alpha}{\lambda}\,\zeta_{\lambda}({\bf p};{\alpha},{\bf n}),
\end{equation}
in the form 
\begin{equation}
\label{10}
\zeta_{\lambda}({\bf p};{\alpha},{\bf n})=A_{\lambda}({\bf p};{\bf n})\,\xi^{(0)}({\bf p};{\alpha},{\bf n}),
\end{equation}
where $({\bf n}=(\sin{\theta}\cos{\varphi},\sin{\theta}\sin{\varphi},\cos{\theta})$
\begin{equation}
\label{11}
\xi^{(0)}({\bf p},{\alpha},{\bf n})=\frac{1}{(2\pi)^{3/2}}[(pn)/m]^{-1+i\alpha},
\end{equation}
are the  eigenfunctions of the operator $C_1$ for the particle with spin zero   \cite{Shapiro,Kad,Ska}).

Substituting (\ref{10}) into equations (\ref{8}) and (\ref{9}), we obtain  for $A_{\lambda}({\bf p};{\bf n})$  
\begin{equation}
\label{12}
C_1\,A_{\lambda}({\bf p};{\bf n})=-\frac{1}{4}A_{\lambda}({\bf p};{\bf n}),\quad{C_2}\,A_{\lambda}({\bf p};{\bf n})=-\imath{\lambda}A_{\lambda}({\bf p};{\bf n}),
\end{equation}
\begin{equation}
\label{13}
[m^{2}{\bf n}{\bf \nabla}_{\bf p}-(pn){\bf p}{\bf \nabla}_{\bf p}-\imath\frac{m{\vec \sigma}\cdot({{\bf p}\times{\bf n}})}{2(p_0+m)}]A_{\lambda}({\bf p};{\bf n})=0,
\end{equation}
\begin{equation}
\label{14}
[m{\vec \sigma}\cdot{\bf n}-\frac{pn+m}{p_0+m}{\vec \sigma}\cdot{\bf p}]A_{\lambda}({\bf p};{\bf n})=2\lambda(pn)A_{\lambda}({\bf p};{\bf n}).
\end{equation}
The calculations give $(p_{-}=p_{1}-\imath{p_{2}})$
\begin{equation}
\label{15}
A_{1/2}({\bf p};{\bf n})=B\,\left( \begin{array}{ccc}(p_0+m-p_3)e^{-i\varphi/2}\cos({\theta}/2)-p_{-}e^{i\varphi/2}\sin({\theta}/2) \\(p_0+m+p_3)e^{i\varphi/2}\sin({\theta}/2)-p_{+}e^{-i\varphi/2}\cos({\theta}/2)\end{array}\right),
\end{equation}
\begin{equation}
\label{16}
A_{-1/2}({\bf p};{\bf n})=B\,\left( \begin{array}{ccc}p_{-}e^{i\varphi/2}\cos({\theta}/2)-(p_0+m+p_3)e^{-i\varphi/2}\sin({\theta}/2) \\(p_0+m-p_3)e^{i\varphi/2}\cos({\theta}/2)-p_{+}e^{-i\varphi/2}\sin({\theta}/2)\end{array}\right),
\end{equation}
where 
\begin{equation}
\label{17}
B({\bf p};{\bf n})=\frac{1}{\sqrt{2(p_0+m)(pn)}}.
\end{equation}
As a result we have  
\begin{equation}
\label{18}
\zeta_{1/2}({\bf p};\alpha,{\bf n})={A}({\bf p};{\bf n})\,\xi^{(0)}({\bf p},\alpha,{\bf n}){\chi}_{1/2},{\quad}\chi_{1/2}=\left( \begin{array}{ccc}1\\0\end{array}\right),
\end{equation}
\begin{equation} 
\label{19}
\zeta_{-1/2}({\bf p};\alpha,{\bf n})={A}({\bf p};{\bf n})\,\xi^{(0)}({\bf p},\alpha,{\bf n}){\chi}_{-1/2},\quad\chi_{-1/2}=\left( \begin{array}{ccc}0\\1\end{array}\right),
\end{equation}
here $A({\bf p};{\bf n})$ is the  matrix whose columns are $A_{1/2}({\bf p};{\bf n})$ and $A_{-1/2}({\bf p};{\bf n})$ 
\begin{equation}
\label{20}
{A}^{\dagger}({\bf p},{\bf n})\,{A}({\bf p},{\bf n})=1.
\end{equation}
The completeness and orthogonality relations for the functions $\zeta_{\lambda}({\bf p};\alpha,{\bf n})$ and $\zeta_{\lambda}^{\ast}({\bf p};\alpha,{\bf n})$ have the form
\begin{eqnarray}
\label{21}
\int({\alpha}^2+1/4)d{\alpha}\,d{\omega}_{\bf n}\,\zeta^{\ast}_{{\lambda}_{1}}({\bf p};\alpha,{\bf n})\,
\zeta_{{\lambda}_2}({\bf p}^{'};\alpha,{\bf n})={\delta}_{{\lambda}_1{\lambda}_2}{p_0}\delta({\bf p}-{\bf p}^{'}),
\end{eqnarray}
\begin{eqnarray}
\label{22}
\sum_{\lambda=-1/2}^{1/2}\int\frac{d{\bf p}}{p_0}\,\zeta^{\ast}_{\lambda}({\bf p};\alpha,{\bf n})\,\zeta_{\lambda}({\bf p};\alpha^{'},{\bf n}^{'})=\frac{\delta({\bf n}-{\bf n}^{'})\,\delta(\alpha-\alpha^{'})}{\alpha^2+1/4}.
\end{eqnarray}
Using in (\ref{6}) the functions $\zeta_{1/2}({\bf p};\alpha,{\bf n})$ and $\zeta_{-1/2}({\bf p};\alpha,{\bf n})$  we obtain for the  transition amplitudes with ${\lambda}={\lambda}^{'}=1/2$ and ${\lambda}={\lambda}^{'}=-1/2$ equal integral representation
\begin{eqnarray}
\label{23}
K(x_2;x_1,1/2,n)&=&{\lefteqn{\int\frac{d{\bf p}}{p_0}\,{\zeta}^{\ast}_{1/2}({\bf p},{\alpha},{\bf n})\,S(x_2-x_1){\zeta}_{1/2}({\bf p},{\alpha},{\bf n})}}\nonumber\\&&=\frac{1}{(2\pi)^{3}}\int\frac{d{\bf p}}{p_0}\frac{\exp-i[(x_2-x_1)p]}{[(pn)/m]^2},
\end{eqnarray} 
\begin{eqnarray}
\label{24}
K(x_2;x_1,-1/2,n)&=&{\lefteqn{\int\frac{d{\bf p}}{p_0}\,{\zeta}^{\ast}_{-1/2}({\bf p},{\alpha},{\bf n})\,S(x_2-x_1){\zeta}_{-1/2}({\bf p},{\alpha},{\bf n})}}\nonumber\\&&=\frac{1}{(2\pi)^{3}}\int\frac{d{\bf p}}{p_0}\frac{\exp-i[(x_2-x_1)p]}{[(pn)/m]^2}.
\end{eqnarray}   
These transition amplitudes contain  the vector  of the  light-cone ${n}$ and have the same form as the transition amplitude for the particle with spin zero.
\subsubsection*{ACKNOWLEDGMENTS}

This work was supported by the Deutsche Forschungsgemeinschaft (No. FR 1560/1-1).

\end{document}